\documentclass[showpacs,twocolumn,prb]{revtex4}
\usepackage{amssymb}
\usepackage{graphicx}

\begin{document}

\title{Current-induced Spin Polarization in Two-Dimensional Hole Gas}
\author{Chao-Xing Liu$^{1,2}$, Bin Zhou$^{2}$, Shun-Qing Shen$^{2}$, and
Bang-fen Zhu$^{1}$ }
\affiliation{$^{1}$Department of Physics and Center for Advanced Study, Tsinghua
University, Beijing 100084, China\\
$^{2}$Department of Physics and Center of Computational and Theoretical
Physics, The University of Hong Kong, Hong Kong, China}
\date{\today}

\begin{abstract}
We investigate the current-induced spin polarization in the two-dimensional
hole gas (2DHG) with the structure inversion asymmetry. By using the
perturbation theory, we re-derive the effective $k$-cubic Rashba Hamiltonian
for 2DHG and the generalized spin operators accordingly. Then based on the
linear response theory we calculate the current-induced spin polarization
both analytically and numerically with the disorder effect considered. We
have found that, quite different from the two-dimensional electron gas, the
spin polarization in 2DHG depends linearly on Fermi energy in the low doping
regime, and with increasing Fermi energy, the spin polarization may be
suppressed and even changes its sign. We predict a pronounced peak of the
spin polarization in 2DHG once the Fermi level is somewhere between minimum
points of two spin-split branches of the lowest light-hole subband. We
discuss the possibility of measurements in experiments as regards the
temperature and the width of quantum wells.
\end{abstract}

\maketitle

\section{Introduction}
In order to reduce the electric leakage and to meet the challenge brought
about by the reduced physical size of the future nano-electronics, it is
being explored to replace the electron charge with the spin degree of
freedom in the electronic transport. This is the ambitious goal of
researchers in the field of spintronics.~\cite{Wolf,Zutic,Awschalom} One of
basic issues in this field is how to generate the polarized spin in devices.
As an straightforward way, the spin injection from ferromagnetic layers may
provide a possible solution to this problem if the interface mismatch
problem can be avoided, but it is more desirable to generate spin
polarization directly by electric means in devices because of its easy
controllability and compatibility with the standard microelectronics
technology.~\cite{Wolf,Zutic,Awschalom} The spin-orbit coupling (SOC) in
semiconductors, which relates the electron spin to its momentum, may provide
a controllable way to realize such purpose. Based on this idea, the
phenomenon of current-induced spin polarization (CISP) has recently
attracted extensive attentions of a lot of research groups.~\cite%
{Dyakonov,Edelstein,Aronov,Chaplik,Inoue1,Bleibaum1,Bleibaum2,Vavilov,
Tarasenko,Trushin,Huang,LiangbinHu,Xiaohua,Bao06,Silov1,Kato1,Kato2,Sih,Stern,
Yang,Ganichev1,Ganichev2,Cui}

As early as in 1970's, the CISP due to the spin-orbit scattering near the
surface of semiconductor thin films was predicted by Dyakonov and Perel.~%
\cite{Dyakonov} Restricted by experimental conditions at that time, this
prediction was ignored until the beginning of 1990's. With the development
of sample fabrication and characterization technology in low-dimensional
semiconductor systems, it was realized that such phenomena could also exist
in quantum wells and heterostructures with the structure or bulk inversion
asymmetry.~\cite{Edelstein,Aronov} Later, many interesting topics about CISP
have been raised, such as the joint effect of the Rashba and Dresselhaus SOC
mechanism,~\cite{Chaplik} vertex correction,~\cite%
{Edelstein,Aronov,Chaplik,Inoue1} quantum correction~\cite%
{Bleibaum1,Bleibaum2} and resonant spin polarization.~\cite{Bao06}
Experimentally, CISP was first observed by Silov \textit{et al}~\cite{Silov1}
in two-dimensional hole gas (2DHG) by using the polarized photoluminescence.~%
\cite{Kaestner1,Kaestner2,Silov2} When inputting an in-plane current into
the 2DHG system, they observed a large optical polarization in
photoluminescence spectra.~\cite{Silov1} Later, Kato \textit{et al}
demonstrated the existence of the CISP in strained nonmagnetic
semiconductors,~\cite{Kato1,Kato2} and Sih \textit{et al} detected the CISP
in the two-dimensional electron gas (2DEG) in (110) $AlGaAs$ quantum well.~%
\cite{Sih} The CISP was also found in $ZnSe$ epilayers even up to the room
temperature.~\cite{Stern} Very recently, the converse effect of CISP has
been clearly shown by Yang \textit{et al} experimentally,\cite{Yang} and the
spin photocurrent has also been observed.~\cite{Ganichev1,Ganichev2,Cui}

So far most theoretic investigations about the CISP deal with the electron
SOC systems.~\cite%
{Dyakonov,Edelstein,Aronov,Chaplik,Inoue1,Bleibaum1,Bleibaum2,Vavilov,Tarasenko,Huang,Xiaohua,Trushin,LiangbinHu}
Thus the CISP in the 2DHG system as shown in Silov's experiments was also
interpreted in terms of the linear-$k$ Rashba coupling of the 2DEG systems
with several parameters adjusted.~\cite{Silov1} As we shall show later, this
treatment is not appropriate for 2DHG. Unlike the electron system, the hole
state in the Luttinger-Kohn Hamiltonian~\cite{Luttinger} is a spinor of four
components. As each component is a combination of spin and orbit momentum,
the spin of a hole spinor is not a conserved physical quantity. Therefore,
the "spintronics" for hole gas is in fact a combination of spintronics and
orbitronics\cite{Bernevig2}. If only the lowest heavy hole (HH1) subband is
concerned, by projecting the multi-band Hamiltonian of 2DHG with structural
inversion asymmetry into a subspace spanned by $|\pm \frac{3}{2}\rangle$
mostly relevant with the HH1 states, we can obtain the $k$-cubic Rashba
model~\cite{Schliemann,Bernevig,Sinova,Winkler1,Winkler2}. We emphasize here
in this lowest heavy hole subspace, the spin operators are no longer
represented by three Pauli matrices, because the "generalized spin" we shall
adopt is a hybridization of spin and orbit angular momentum. In deriving the
effective Hamiltonian from the Luttinger-Kohn Hamiltonian by the
perturbation and truncation procedure to higher orders, one must take care
of the corresponding transformation for the spin operator in order to obtain
the correct expression. In the following, we will use the terminology
"generalized spin", or the "spin" for short, to denote the total angular
momentum in the spin-orbit coupled systems.

The aim of the present paper is to investigate the CISP of 2DHG in a more
rigorous way. Namely, we will derive the $k$-cubic Rashba model and the
corresponding spin operators for holes, and on this basis we will present
both analytical and numerical results for the CISP in 2DHG. This paper is
organized as follows. In Sec II the general formalism and the Hamiltonian
for the 2DHG with structural inversion asymmetry is given. In Sec III in the
low doping regime, with the perturbation theory, the Hamiltonian and spin
operators in the lowest heavy hole subspace are derived, and applied to
analytical calculation of the CISP in 2DHG.
In Sec IV, we will show the numerical calculations agree well with the
analytical results at the low-doping regime; while in the high doping regime
the numerical results predict some new features of CISP. Particularly, we
predict a pronounced CISP peak when Fermi energy lies little above the
energy minimum of the lowest light hole (LH1) subband. Finally, a brief
summary is drawn.

\section{Formalism}

\subsection{Hole Hamlitonian}

A p-doped quantum well system with structural inversion asymmetry can be
described as the isotropic Luttinger-Kohn Hamiltonian with a confining
asymmetrical potential,
\begin{eqnarray}
\hat{H}=\hat{H}_{L}+ \hat{V}_{c}(z)+ \hat{V}_{a}(z).  \label{Ham1}
\end{eqnarray}%
Here in order to compare the analytical results with the numerical one, the
confining potential along the z-direction $V_{c}(z)$ is taken as
\begin{eqnarray}
\hat{V}_{c}(z) = & \left \{
\begin{array}{cc}
0 & -L_z /2 < z < L_z /2 \\
\infty & otherwise,%
\end{array}
\right.  \label{Vc}
\end{eqnarray}
where $L_z$ is the well width of the quantum well. The asymmetrical
potential, which stems from a build-in electric field $F$ via the gate
voltage or $\delta $-doping is $\hat{V}_{a}(z)=eFz,$ which breaks the
inversion symmetry and lifts the spin doublet degeneracy.

Let $\hat{S}$ be the generalized spin operator of a hole state, and $\hat{S}%
_{z}$ be the z-component of $\hat{S}$, the isotropic Luttinger-Kohn
Hamiltonian $\hat{H}_{L}$ in the $\left\vert S, S_{z}\right\rangle $
representation (four basis kets written in the sequence of $\{|\frac{3}{2}%
\rangle ,|\frac{1}{2}\rangle ,|-\frac{1}{2}\rangle,|-\frac{3}{2}\rangle \}$)
is expressed as
\begin{equation}
\hat{H}_{L}=\left(
\begin{array}{cccc}
P & R & T & 0 \\
R^{\dag } & Q & 0 & T \\
T^{\dag } & 0 & Q & -R \\
0 & T^{\dag } & -R^{\dag } & P%
\end{array}%
\right) ,  \label{HamLut}
\end{equation}%
with
\begin{eqnarray}
P &=&\frac{\hbar ^{2}}{2m_{0}}[(\gamma _{1}+\gamma _{2})\mathbf{k}
^{2}+(\gamma _{1}-2\gamma _{2})k_{z}^{2}], \\
Q &=&\frac{\hbar ^{2}}{2m_{0}}[(\gamma _{1}-\gamma _{2})\mathbf{k}
^{2}+(\gamma _{1}+2\gamma _{2})k_{z}^{2}], \\
R &=&-\frac{\hbar ^{2}\sqrt{3}\gamma _{2}}{m_{0}}\mathbf{k}_{-}k_{z}, \\
T &=&-\frac{\hbar ^{2}\sqrt{3}\gamma _{2}}{2m_{0}}\mathbf{k}_{-}^{2},
\end{eqnarray}%
where $\gamma _{1},\gamma _{2}$ is the Luttinger parameters, $m_{0}$ is the
free electron mass, the in-plane wave vector $\mathbf{k}=(k_{x},k_{y})$,
denoted in the polar coordinate as $\mathbf{k }\equiv (k,\theta )$, $\mathbf{%
k}_{\pm}\equiv k_x \pm i k_y$ and $k_{z}=-i\partial /\partial z$. The other
terms, such as anisotropic term, C terms or hole Rashba term,~\cite%
{Bfzhu,Winkler2,Winkler1} have only negligible effects and are omitted in our
calculation. Correspondingly, the $x$-, $y$-, $z$- component of the "spin"-$%
\frac{3}{2}$ operator respectively reads
\begin{eqnarray}
\hat{S}_{x} &=&\frac{1}{2}\left(
\begin{array}{cccc}
0 & \sqrt{3} & 0 & 0 \\
\sqrt{3} & 0 & 2 & 0 \\
0 & 2 & 0 & \sqrt{3} \\
0 & 0 & \sqrt{3} & 0%
\end{array}%
\right) ,  \label{Sx0}
\end{eqnarray}%
\begin{eqnarray}
\hat{S}_{y}&=&\frac{i}{2}\left(
\begin{array}{cccc}
0 & -\sqrt{3} & 0 & 0 \\
\sqrt{3} & 0 & -2 & 0 \\
0 & 2 & 0 & -\sqrt{3} \\
0 & 0 & \sqrt{3} & 0%
\end{array}%
\right) ,  \label{Sy0}
\end{eqnarray}%
\begin{eqnarray}
\hat{S}_{z}&=&\frac{1}{2}\left(
\begin{array}{cccc}
3 & 0 & 0 & 0 \\
0 & 1 & 0 & 0 \\
0 & 0 & -1 & 0 \\
0 & 0 & 0 & -3%
\end{array}%
\right) .  \label{Sz0}
\end{eqnarray}
We stress here again that the "spin" of the $\frac{3}{2}$ spinor is actually
its total angular momentum, which is a linear combination of spin and orbit
angular momentum of a valence band electron. In polarized optical
experiments, such as polarized photoluminescence~\cite%
{Kaestner1,Kaestner2,Silov2} or Kerr/Farady rotation~\cite{Kato1,Kato2}, it
is appropriate to introduce such a generalized spin.

For the infinitely confining potential, we expand the eigenfunction $\phi
_{\nu }$ associated with the $\nu th$ hole subband in terms of confined
standing waves as
\begin{equation}
\phi _{\nu }(\mathbf{k})=\sum_{n,\lambda _{h}}a_{n,\lambda _{h}}^{\nu }(%
\mathbf{k})\frac{1}{2\pi }e^{i\mathbf{k}\cdot \mathbf{r}}|n,\lambda
_{h}\rangle _{h},  \label{Basis}
\end{equation}%
with
\begin{equation}
|n,\lambda _{h}\rangle =\sqrt{\frac{2}{L_z}}\sin \left( \frac{n\pi (z+ L_z/2)%
}{L_z}\right) |\lambda _{h}\rangle ,  \label{Basiswf}
\end{equation}%
where $\mathbf{r}=(x,y)$, $n$ is the confinement quantum number for the
standing wave along the $z$-direction, and $\lambda _{h}$ denotes the $%
\lambda _{h}$-component of the hole ($\lambda _{h}=3/2,1/2,-1/2,-3/2$).
Since we are only interested in the low energy physics, a finite number of $%
n $ will result in a reasonable accuracy, and the effective Hamiltonian is
reduced into a square matrix with a dimension of $4n$. In this way we obtain
the hole subband structure analytically or numerically.

\subsection{Expression for CISP}

In the framework of the linear response theory, the electric response of
spin polarization in a weak external electric field $\mathbf{E}$ can be
formulated as~\cite{Bao06}
\begin{equation}
\langle \hat{S}_{\alpha }\rangle=\sum_{\beta }\chi _{\alpha \beta }E_{\beta
},  \label{Spinsus1}
\end{equation}%
where $\langle \hat{S}_{\alpha }\rangle$ is the thermodynamically averaged
value of the spin density. The electric spin susceptibility $\chi _{\alpha
\beta}$ can be calculated by Kubo formula.~\cite{Mahan} By the Green
function formalism, the Bastin version of Kubo formula~\cite{Streda} reads
\begin{equation}
\chi _{\alpha \beta }=\frac{ie\hbar }{2\pi }\int dEf(E)\text{Tr}\left\langle
\hat{S}_{\alpha }\left( \frac{dG^{R}}{dE}v_{\beta }A-Av_{\beta } \frac{dG^{A}%
}{dE}\right)\right\rangle _{c},  \label{Kubo1}
\end{equation}%
where $G^{R}$ and $G^{A}$ are the retarded and advanced Green function,
respectively, $A=i(G^{R}-G^{A})$ is the spectral function, $f(E)$ is the
Fermi distribution function, $v_{\beta }$ is the velocity operator along the
$\beta$ direction, and the bracket $\langle \cdots \rangle _{c}$ represents
the average over the impurity configuration.

To taken the vertex correction into account, we use the Streda-Smrcka
division of Kubo formula,~\cite{Streda,Sinitsyn}
\begin{equation}
\chi _{\alpha \beta }=-\frac{e\hbar }{2\pi }\int dE\frac{\partial f(E)}{%
\partial E}Tr\langle \hat{S}_{\alpha }G^{R}(E_{F})v_{\beta
}G^{A}(E_{F})\rangle _{c},  \label{Kubo2}
\end{equation}%
in which we retain only the non-analytical part, and neglect the analytical
part, because the latter is much less important in the present case. In the
following, we will use Eq.~(\ref{Kubo2}) to analytically calculate the
electric spin susceptibility (ESS) with the vertex correction considered;
meanwhile we will carry out the numerical calculation with Eq.~(\ref{Kubo1})
in the relaxation time approximation. We shall show that the analytical and
numerical results are in good agreements with each other in the regime of
low hole density.

\subsection{Symmetry}

The general properties of $\chi _{\alpha \beta }$ will be critically
determined by symmetry of the system. For the two-dimensional system we
investigate, the index $\alpha $($\beta $) in Eq.~(\ref{Spinsus1}) is simply
chosen to be $x$ or $y$ in the following. Without the asymmetrical potential
$V_{a}$, the Hamiltonian (\ref{Ham1}) is invariant under the space inversion
transformation
\begin{equation}
\begin{array}{ccc}
x\rightarrow -x, & y\rightarrow -y, & z\rightarrow -z, \\
\hat{S}_{x}\rightarrow \hat{S}_{x}, & \hat{S}_{y}\rightarrow \hat{S}_{y}, &
\hat{S}_{z}\rightarrow \hat{S}_{z},%
\end{array}
\label{Traninv}
\end{equation}%
if the origin point of $z$-axis is set at the mid-plane of the quantum well.
Applying the space inversion transformation (\ref{Traninv}) to Eq.~(\ref%
{Spinsus1}), we have
\begin{equation}
\langle \hat{S}_{\alpha }\rangle=\chi _{\alpha \beta }E_{\beta }\rightarrow
\langle \hat{S}_{\alpha }\rangle=-\chi _{\alpha \beta }E_{\beta },
\end{equation}%
whereby $\chi _{\alpha \beta }=-\chi _{\alpha \beta }$. This implies that no
CISP appears when the inversion symmetry exists in the system. So the
asymmetrical potential $V_{a}$ is crucial for the CISP.

In the presence of an asymmetrical potential $V_{a}$, the Hamiltonian (\ref%
{Ham1}) is invariant versus the rotation along z-axis with $\frac{\pi }{2}$
in both the real space and the spin space,
\begin{equation}
\begin{array}{ccc}
x\rightarrow y, & y\rightarrow -x, & z\rightarrow z, \\
\hat{S}_{x}\rightarrow \hat{S}_{y}, & \hat{S}_{y}\rightarrow -\hat{S}_{x}, &
\hat{S}_{z}\rightarrow \hat{S}_{z}.%
\end{array}
\label{Tran1}
\end{equation}%
With the above transformations (\ref{Tran1}), Eq.~(\ref{Spinsus1}) will give
\begin{eqnarray}
\langle \hat{S}_{x}\rangle &=&\chi _{xy}E_{y}\rightarrow \langle \hat{S}%
_{y}\rangle=-\chi _{xy}E_{x}, \\
\langle \hat{S}_{x}\rangle &=&\chi _{xx}E_{x}\rightarrow \langle \hat{S}%
_{y}\rangle=\chi _{xx}E_{y}.
\end{eqnarray}%
Combined with $\langle \hat{S}_{y}\rangle=\chi _{yx}E_{x}$ and $\langle \hat{%
S}_{y}\rangle =\chi _{yy}E_{y}$, we get
\begin{eqnarray}
\chi _{xy} &=&-\chi _{yx},  \label{chi1} \\
\chi _{xx} &=&\chi _{yy},  \label{chi2}
\end{eqnarray}%
which are direct consequence of the rotation symmetry along the z-axis.

\section{Analytical Results for CISP in 2DHG}

In the low hole density regime an effective Hamiltonian can be obtained by
projecting the Hamiltonian (\ref{Ham1}) into the subspace spanned by the
lowest heavy hole states, which, by using the truncation approximation and
projection perturbation method, \cite{Bfzhu,Shen00prb,Winkler1,Winkler2,Foreman,Foreman2,Foreman3,Habib}
 is reduced to the widely
used $k$-cubic Rashba model. More importantly, the corresponding spin
operators in the subspace will be obtained properly, and the ESS of 2DHG
with the impurity vertex correction will be worked out. Then we will compare
and contrast the different behaviors of the CISP in the 2DEG and 2DHG in
this Section.

\subsection{ $k$-cubic Rashba Model}

To obtain an approximate analytical expression, we take the following
procedure. First we expand a hole state in terms of 8 basis wave functions
associated with $|n,\lambda _{h}\rangle $ ($n=1,2$ and $\lambda _{h} = \frac{%
3}{2}, \frac{1}{2}, -\frac{1}{2}, -\frac{3}{2}$) ( Eq.~\ref{Basiswf}). Then
for a given $\mathbf{k}$, we may express the Hamiltonian (\ref{Ham1}) in
terms of an $8\times 8$ matrix, which by the perturbation procedure can be
further projected into the subspace spanned by the $|1, \frac{3}{2}\rangle$
and $|1, -\frac{3}{2}\rangle$ states. Thus we obtain a $2\times 2$ matrix as
( See Appendix A for details),
\begin{equation}
\hat{H}_{k^{3}}=\frac{\hbar ^{2}k^{2}}{2m_{h}}+i\alpha (k_{-}^{3}\sigma
_{+}-k_{+}^{3}\sigma _{-}),  \label{HamkR3}
\end{equation}%
where the Pauli matrix $\sigma _{\pm }\equiv \frac{1}{2}(\sigma _{x}\pm i
\sigma _{y})$, %$k _{\pm } \equiv k_{x}\pm i k_{y}$,
the effective mass is renormalized into
\begin{equation}
m_{h}=m_{0}\left( \gamma _{1}+\gamma _{2}-\frac{256\gamma _{2}^{2}}{3\pi
^{2}(3\gamma _{1}+10\gamma _{2})}\right) ^{-1},
\end{equation}%
and the $k$-cubic Rashba coefficient
\begin{equation}
\alpha =\frac{512eFL_{z}^{4}\gamma _{2}^{2}}{9\pi ^{6}(3\gamma _{1}+10\gamma
_{2})(\gamma _{1}-2\gamma _{2})}.
\end{equation}%
Note that Eq.~(\ref{HamkR3}) is just the $k$-cubic Rashba model, in which
the Rashba coefficient $\alpha$ is proportional to asymmetrical potential
strength $F$, in agreement with the results by Winkler.~\cite{Winkler1} We
can rewrite the $k$-cubic Rashba Hamiltonian (\ref{HamkR3}) as
\begin{equation}
\hat{H}_{k^{3}}=\varepsilon (\mathbf{k})+\sum_{i=x,y,z}d_{i}(\mathbf{k}%
)\sigma _{i},
\end{equation}%
where $d_{x}=\alpha k_{y}(3k_{x}^{2}-k_{y}^{2})$, $d_{y}=\alpha k_{x}
(3k_{y}^{2}-k_{x}^{2})$, $d_{z}=0$, and $\varepsilon (\mathbf{k})=\frac{%
\hbar ^{2}k^{2}}{2m_{h}}$. The eigenvalue associated with the spin index $%
\mu $ ($\mu =\pm 1$) is
\begin{equation}
E_{\mu }(k)=\varepsilon (\mathbf{k})+\mu \alpha k^{3},
\end{equation}%
with the eigenfunction
\begin{equation}
\psi _{\mathbf{k}\mu }(\mathbf{r})=\frac{e^{i\mathbf{k}\cdot \mathbf{r}}}{%
\sqrt{2A_S}}\left(
\begin{array}{c}
i \\
\mu e^{i3\theta }%
\end{array}%
\right) ,
\end{equation}%
where $A_S$ is the area of the system.

The $k$-cubic Rashba model has been widely used to study the spin Hall
effect in 2DHG;~\cite{Schliemann,Bernevig,Sinova} however, no sufficient
attention has been paid to the corresponding spin operators. For example,
although Hamiltonian (\ref{HamkR3}) is written in terms of the Pauli
matrices $\sigma $, the $\sigma $ matrix is no longer related to the spin
directly. The correct spin operators in the $k$-cubic Rashba model, as
described in Appendix A, are expressed as
\begin{eqnarray}
\tilde{S}_{x} &=&\left(
\begin{array}{cc}
-S_{0}k_{y} & S_{1}k_{-}^{2} \\
S_{1}k_{+}^{2} & -S_{0}k_{y}%
\end{array}%
\right) ,  \label{Sx1} \\
\tilde{S}_{y} &=&\left(
\begin{array}{cc}
S_{0}k_{x} & -iS_{1}k_{-}^{2} \\
iS_{1}k_{+}^{2} & S_{0}k_{x}%
\end{array}%
\right) ,  \label{Sy1} \\
\tilde{S}_{z} &=&\frac{3}{2}\sigma _{z},  \label{Sz1}
\end{eqnarray}%
in which
\begin{eqnarray}
S_{0} &=&\frac{512\gamma _{2}L_{z}^{4}eFm_{0}}{9\pi ^{6}\hbar ^{2}(3\gamma
_{1}+10\gamma _{2})(\gamma _{1}-2\gamma _{2})},  \label{S0kR3} \\
S_{1} &=&\left[ \frac{3}{4\pi ^{2}}-\frac{256\gamma _{2}^{2}}{3\pi
^{4}(3\gamma _{1}+10\gamma _{2})^{2}}\right] L_{z}^{2}.  \label{S1kR3}
\end{eqnarray}%
Clearly, the coefficient $S_{0}$ and the Rashba coefficient $\alpha$ have
the same dependence on $F$ and $L_{z}$, thus we have
\begin{equation}
S_{0}=\frac{\alpha m_{0}}{\hbar^{2}\gamma _{2}}.  \label{S0alphakR3}
\end{equation}%
$S_{z}$ is related to $\sigma _{z}$, while $S_{x}(S_{y})$ consists of two
parts: the diagonal part linear in $k_{y}(k_x)$ and the non-diagonal part
quadratic in $k_{\pm }$. The diagonal part, which relates the wave vector $%
k_{y}$ ($k_{x})$ with $S_{x}$ ($S_{y}),$ will give the main contribution to
CISP.
%It is noticed that this relation above can be utilized to deduce the $k$-cubic Rashba coefficient from the CISP.
The velocity operator in the $k$-cubic Rashba model can also be obtained by
the projection technique,
\begin{equation}
\tilde{v}_{x}=\frac{\hbar k_{x}}{m_{h}}+\frac{3i\alpha }{\hbar }%
(k_{-}^{2}\sigma _{+}-k_{+}^{2}\sigma _{-}),  \label{VxkR3}
\end{equation}%
which is consistent with the relation $\tilde{v}_{x}=\frac{1}{\hbar }%
\partial H_{k^{3}}/\partial k_{x}$.

\subsection{Impurity Vertex correction}

Now, we calculate the ESS in the framework of the linear response theory
based on $k$-cubic Rashba model (\ref{HamkR3}). In doing this we take the
vertex correction of impurities into account. The free retarded Green
function has the form,
\begin{equation}
G_{0}^{R}(\mathbf{k},E)=\frac{E-\varepsilon (\mathbf{k})+
\sum_{i}d_{i}\sigma _{i}}{(E-E_{+}+i\eta )(E-E_{-}+i\eta )},
\end{equation}%
where $\eta $ is an infinitesimal positive number. We assume impurities to
be distributed randomly in the form $V_{r}(\mathbf{r})=V_{0}\sum_{i}\delta(%
\mathbf{r}-\mathbf{R}_{i})$, where $V_{0}$ is the strength. With the Born
approximation, the self-energy, diagonal in the spin space, is given by
\begin{equation}
Im[\Sigma _{0}^{R}(\mathbf{k},E)]=\frac{n_{i}V_{0}^{2}\pi }{2}(D_{+}+D_{-}),
\end{equation}%
where $n_{i}$ is the impurity density, and the density of states for two
spin-split branches of the HH1 subband reads
\begin{equation}
D_{\pm }(k)=\frac{m_h}{2\pi \hbar ^{2}}\left|1\pm \frac{3m_h \alpha k}{\hbar
^{2}}\right|^{-1}.
\end{equation}%
So the configuration-averaged Green function is given by
\begin{equation}
G^{R}(\mathbf{k},E)=\frac{E-\varepsilon (k)+i\Gamma _{0}+\sum_{i}d_{i}\sigma
_{i}}{(E-E_{+}+i\Gamma _{0})(E-E_{-}+i\Gamma _{0})},
\end{equation}%
where $\Gamma _{0}=-Im[\Sigma _{0}^{R}(\mathbf{k},E)]=\frac{\hbar }{2\tau }$
and $\tau $ is the momentum relaxation time. In the ladder approximation,
the Strda-Smrcka formula (\ref{Kubo2}) for the ESS $\chi$ will reduce to
\begin{equation}
\chi _{\alpha\beta}=e\hbar \int \frac{dE}{2\pi }\left( -\frac{\partial f(E)}{%
\partial E}\right) \int \frac{d^{2}k}{(2\pi )^{2}}Tr\left[ \tilde{S}%
_{\alpha}G^{R}\Upsilon _{\beta}G^{A}\right] ,  \label{Kubo3}
\end{equation}%
where $\mathbf{\tilde{S}}$ is given by Eqs.(\ref{Sx1})-(\ref{Sz1}) and the
vertex function $\Upsilon _{\beta}(\mathbf{k})$ satisfies the
self-consistent equation~\cite{Mahan}
\begin{equation}
\Upsilon _{\beta}=\tilde{v}_{\beta}+n_{i}V_{0}^{2}\int \frac{d^{2}k}{(2\pi
)^{2}}G^{R}(\mathbf{k},E)\Upsilon _{\beta}G^{A}(\mathbf{k},E).
\label{VereqnkR3}
\end{equation}%
Suppose the electric field is along the x-direction, we solve the vertex
function $\Upsilon_{x}$ iteratively, and get the first-order correction to $%
\Upsilon _{x} $ as
\begin{widetext}
\begin{eqnarray}
\Delta\Upsilon_x^{(1)}&=&n_iV_0^2\int\frac{kdkd\theta}{(2\pi)^2}\frac{\left(\begin{array}{cc}E-\varepsilon(\bold{k})&i\alpha
k_-^3\\-i\alpha k_+^3&E-\varepsilon(\bold{k})\end{array}\right)\left(\begin{array}{cc}\frac{\hbar k_x}{m_h}&\frac{3i\alpha}{\hbar}k_-^2\\
\frac{-3i\alpha}{\hbar}k_+^2&\frac{\hbar
k_x}{m_h}\end{array}\right)\left(\begin{array}{cc}E-\varepsilon(\bold{k})&i\alpha
k_-^3\\-i\alpha k_+^3&E-\varepsilon(\bold{k})\end{array}\right)}
{((E-E_+)^2+\Gamma_0^2)((E-E_-)^2+\Gamma_0^2)}.\label{Vereqn1kR3}
\end{eqnarray}
\end{widetext}
Note that $E_{\pm}$ and $\Gamma _{0}$ are independent of $\theta $ and all
terms in the numerator of the integrand contain something like $exp(\pm
i\theta)$ etc., so the integral over $\theta $ from $0$ to $2\pi $ in Eq.(%
\ref{Vereqn1kR3}) vanishes. Furthermore, the higher order terms for the
vertex correction vanish either, which is quite different from the vertex
correction in the linear-$\mathbf{k}$ Rashba model.~\cite{Inoue1} The same
situation occurs for $\Upsilon_y$. The above results agree with the work by
Schliemann and Loss.~\cite{Schliemann} The calculation of the spin
polarization is straightforward, and to the lowest order in Fermi momentum $%
k_{\pm }^{F}$ and $\alpha$, only the term proportional to $S_{0}$
contributes to the spin polarization. The final result reads
\begin{eqnarray}
\chi _{yx}&=&-\chi_{xy}=\frac{eS_{0}\tau m_{h}E_{F}}{\hbar ^{3}\pi }%
=S_{0}n_{h}\frac{e\tau }{\hbar },  \label{ChiyxkR3} \\
\chi_{xx}&=&\chi_{yy}=0,  \label{ChixxkR3}
\end{eqnarray}
where $n_{h}$ is the hole density, $E_{F}$ is the Fermi energy, and only the
leading term in $E_F$ is retained.

In the relaxation time approximation the longitudinal conductivity of 2DHG
equals to
\begin{equation}
\sigma _{xx}=\frac{e^{2}\tau E_{F}}{\hbar ^{2}\pi }.  \label{ConducxxkR3}
\end{equation}%
Thus, combining the expressions (\ref{ChiyxkR3}) and (\ref{ConducxxkR3}), we
have the ratio
\begin{equation}
\frac{\langle \tilde{S}_{y}\rangle }{\langle j_{x}\rangle }=\frac{\chi _{yx}%
}{\sigma _{xx}}=\frac{S_{0}m_{h}}{e\hbar }=\frac{\alpha m_{0}m_{h}}{e\hbar
^{3}\gamma _{2}}.  \label{SyjxkR3}
\end{equation}%

The formula above can be also obtained from the expression of the
spin operator (\ref{Sy1}) and the velocity operator (\ref{VxkR3}) by
neglecting the non-diagonal part in the spin operator and the
anomalous part in the velocity operator, i.e. $\tilde{S}_y \approx
S_0k_x$ and $j_x \approx{e\hbar k_x}/{m_h}$. Obviously, this ratio
depends only on the material parameters, but not on the impurity
scattering nor the carrier density in the low density limit.
Meanwhile, since both the current and spin polarization can be
measured experimentally, the relation (\ref{SyjxkR3} ) may be
invoked to obtain the $k$-cubic Rashba coefficient $\alpha $
experimentally.

\subsection{ Comparing CISP of 2DHG and 2DEG}

The CISP of 2DHG manifests itselve several features different from that of
2DEG. To illustrate this, let's first take a look at the CISP of 2DEG. The
electric spin susceptibility is given by $\chi _{yx}={2e\tau \alpha _{e}m_{e}%
}/\hbar ^{2}$, where $m_{e}$ is the effective mass of electron and $\alpha
_{e}$ is the linear Rashba coefficient. As shown by Inoue \textit{et al.}%
\cite{Inoue1}, the vertex correction due to the linear Rashba spin splitting
is non-trivial. With the longitudinal conductivity of 2DEG $\sigma _{xx}={%
e^{2}\tau E_{F}}/(\hbar ^{2}\pi )$, we find the ratio of spin polarization
to the current for the 2DEG is
\begin{equation}
\frac{\langle S_{y}^{(e)}\rangle }{\langle j_{x}\rangle }=\frac{\chi_{yx}}{%
\sigma_{xx}}=\frac{2\pi m_{e}\alpha_{e}}{eE_{F}}.
\end{equation}%
Compared with (\ref{SyjxkR3}), we find the CISP of 2DEG is inversely
proportional to Fermi energy. This means the ratio for 2DEG decreases for
heavier doping. This different Fermi-energy dependence stems from the
different types of spin orientation for 2DEG and 2DHG.

The spin orientation, which is the expectation value of spin operator $%
\mathbf{S}$ for an eigenstate, is given by
\begin{eqnarray}
&&\langle k\mu|\tilde{S}_{x}|k\mu\rangle = -S_{0}k\sin \theta +\mu
k^{2}S_{1}\sin \theta ,  \label{Sx3} \\
&&\langle k\mu|\tilde{S}_{y}|k\mu\rangle =S_{0}k\cos \theta -\mu
k^{2}S_{1}\cos \theta ,  \label{Sy3} \\
&&\langle k\mu|\tilde{S}_{z}|k\mu\rangle= 0,  \label{Sz3}
\end{eqnarray}%
for 2DHG, and
\begin{eqnarray}
\langle k\mu| S^{(e)}_{x}|k\mu\rangle &=&-\mu \sin \theta ,  \label{Sxe} \\
\langle k\mu| S^{(e)}_{y}|k\mu\rangle &=&\mu \cos \theta ,  \label{Sye} \\
\langle k\mu| S^{(e)}_{z}|k\mu\rangle &=&0,  \label{Sze}
\end{eqnarray}%
for 2DEG. In the following, we take $\langle \mathbf{S}\rangle_{k\mu}$ as
short for the spin orientation above. Eqs.~(\ref{Sxe}) and (\ref{Sye}) show
that spin orientation for 2DEG depends on the spin index $\mu$, which has
opposite values for the two spin-splitting states. But for 2DHG, the first
term in Eqs.~(\ref{Sx3}) and (\ref{Sy3}) is independent of the spin index $%
\mu$. Hence, when $k$ is small, this spin-index-independent term will
dominate over the $k^2$-term, leading to the same spin orientation for the
hole state with opposite $\mu$. This is quite different from the electron
case. An interesting question may be raised: why the holes with opposite $%
\mu $ have the same spin orientation? In the following, we will
analyze this problem and try to find the origin of this particular
spin orientation for 2DHG.

Let's first have a look at the electron case. Due to the spin-orbit coupling
and inversion asymmetry, two-fold degeneracy of a subband is lifted. For a
given $k$, we denote two spin-split states as $|+\rangle=\cos\frac{\theta}{2}%
e^{-i\phi}|\frac{1}{2}\rangle_z+\sin\frac{\theta}{2} |-\frac{1}{2}\rangle_z$
and $|-\rangle= -\sin\frac{\theta}{2}e^{-i\phi}|\frac{1}{2}\rangle_z +\cos%
\frac{\theta}{2}|-\frac{1}{2}\rangle_z$, where $|\pm\frac{1}{2}\rangle_z$
are the eigenstates of $\sigma_z$. It is easy to verify that $|+\rangle$ and
$|-\rangle$ have the opposite spin orientation, namely $\langle +|\vec{\sigma%
}|+\rangle=-\langle -|\vec{\sigma}|-\rangle$.

Similar to 2DEG, two spin-split hole states in the subspace $|\pm\frac{3}{2}%
\rangle$ can be constructed as $|+\rangle=\cos\frac{\theta}{2}e^{-i\phi}|%
\frac{3}{2}\rangle+\sin\frac{\theta}{2} |-\frac{3}{2}\rangle$ and $%
|-\rangle=-\sin\frac{\theta}{2}e^{-i\phi}|\frac{3}{2}\rangle +\cos\frac{%
\theta}{2}|-\frac{3}{2}\rangle$. By Eqs.~(\ref{Sx0}) and (\ref{Sy0}), we can
verify the matrix elements of $\hat{S}_x$ and $\hat{S}_y$ between $|\frac{3}{%
2}\rangle$ and $|-\frac{3}{2}\rangle$ vanish, and $\langle\pm|\hat{S}%
_x|\pm\rangle = \langle\pm|\hat{S}_y|\pm\rangle=0$. This indicates that in
the subspace $|\pm\frac{3}{2} \rangle$, any superposition of $|\pm\frac{3}{2}%
\rangle$ will not give rise to the spin orientation along the x- or
y-direction. Thus it is necessary to take the higher order perturbation into
account, in particular the perturbation from coupling between $|\pm\frac{3}{2%
}\rangle$ and $|\pm\frac{1}{2}\rangle$.

Now we give the outline on the origin of the hole spin orientation by the
perturbation procedure ( more systematic method can be found in Appendix A).
Suppose the $HH1\pm$ states $\Psi_{hh,\pm}$ can be expanded as
\begin{eqnarray}
\Psi_{hh,\pm}=\Psi^{(0)}_{hh,\pm}+\Psi^{(1)}_{hh,\pm}+\Psi^{(2)}_{hh,\pm}+%
\cdots\cdots,  \label{Psi1}
\end{eqnarray}
where $\Psi^{(i)}_{hh,\pm}$ denotes the $i$th-order perturbed wave function.
With the basis $|n,\lambda_h\rangle$ [Eq. (\ref{Basiswf})] and the $0$%
th-order term
\begin{eqnarray}
\Psi^{(0)}_{hh,\pm}=|1,\pm\frac{3}{2}\rangle,  \label{Psi0}
\end{eqnarray}
we have the first-order correction as
\begin{eqnarray}
&&\Psi^{(1)}_{hh,+}=\frac{|2,\frac{1}{2}\rangle\langle 2,\frac{1}{2}%
|R^{\dag}|1,\frac{3}{2}\rangle } {E_{1,\frac{3}{2}}-E_{2,\frac{1}{2}}}+\frac{%
|1,-\frac{1}{2}\rangle\langle 1,-\frac{1}{2}|T^{\dag}|1,\frac{3}{2}\rangle
} {E_{1,\frac{3}{2}}-E_{1,-\frac{1}{2}}}  \nonumber \\
&&+\frac{|2,\frac{3}{2}\rangle\langle 2,\frac{3}{2}|V_a|1,\frac{3}{2}\rangle%
}{E_{1,\frac{3}{2}}-E_{2,\frac{3}{2}}},  \label{Psi1+}
\end{eqnarray}
\begin{eqnarray}
&&\Psi^{(1)}_{hh,-}=\frac{|1,\frac{1}{2}\rangle\langle 1,\frac{1}{2}|T|1,-%
\frac{3}{2}\rangle } {E_{1,-\frac{3}{2}}-E_{1,\frac{1}{2}}}-\frac{|2,-\frac{1%
}{2}\rangle\langle 2,-\frac{1}{2}|R|1,-\frac{3}{2}\rangle } {E_{1,-\frac{3}{2%
}}-E_{2,-\frac{1}{2}}}  \nonumber \\
&&+\frac{|2,-\frac{3}{2}\rangle\langle 2,-\frac{3}{2}|V_a|1,-\frac{3}{2}%
\rangle}{E_{1,-\frac{3}{2}}-E_{2,-\frac{3}{2}}},  \label{Psi1-}
\end{eqnarray}
and the second-order correction reads
\begin{eqnarray}
&&\Psi^{(2)}_{hh,+}=\frac{|1,\frac{1}{2}\rangle\langle 1,\frac{1}{2}|V_a|2,%
\frac{1}{2}\rangle \langle 2,\frac{1}{2}|R^{\dag}|1,\frac{3}{2}\rangle} {%
(E_{1,\frac{3}{2}}-E_{1,\frac{1}{2}})(E_{1,\frac{3}{2}}-E_{2,\frac{1}{2}})}
\nonumber \\
&&+\frac{|1,\frac{1}{2}\rangle\langle 1,\frac{1}{2}|R^{\dag}|2, \frac{3}{2}%
\rangle \langle 2,\frac{3}{2}|V_a|1,\frac{3}{2}\rangle}{(E_{1,\frac{3}{2}%
}-E_{1,\frac{1}{2}}) (E_{1,\frac{3}{2}}-E_{2,\frac{3}{2}})}+\cdots,
\label{Psi2+}
\end{eqnarray}
and
\begin{eqnarray}
&&\Psi^{(2)}_{hh,-}=-\frac{|1,-\frac{1}{2}\rangle\langle 1,-\frac{1}{2}%
|V_a|2,-\frac{1}{2}\rangle \langle 2,-\frac{1}{2}|R|1,-\frac{3}{2}\rangle} {%
(E_{1,-\frac{3}{2}}-E_{1,-\frac{1}{2}})(E_{1,-\frac{3}{2}}-E_{2,-\frac{1}{2}%
})}  \nonumber \\
&&-\frac{|1,-\frac{1}{2}\rangle\langle 1,-\frac{1}{2}|R|2,-\frac{3}{2}%
\rangle \langle 2,-\frac{3}{2}|V_a|1,-\frac{3}{2}\rangle}{(E_{1,-\frac{3}{2}%
}-E_{1,-\frac{1}{2}}) (E_{1,-\frac{3}{2}}-E_{2,-\frac{3}{2}})}+\cdots.
\label{Psi2-}
\end{eqnarray}
Here $E_{n,\lambda_h}$ stands for the eigenenergy of the state $%
|n,\lambda_h\rangle$. From Eqs.~(\ref{Sx0}) and (\ref{Sy0}), we can see when
$n=1$ the only nonvanishing terms are $\langle 1,\frac{3}{2}|\hat{S}%
_{x(y)}|1,\frac{1}{2} \rangle$ and $\langle 1,-\frac{3}{2}|\hat{S}_{x(y)}|1,-%
\frac{1}{2}\rangle$. Up to the second-order perturbation, two types of terms
can contribute to $\langle\Psi_{hh,\pm}|\hat{S}_{x(y)}| \Psi_{hh,\pm}\rangle$%
.

The first type stems from the first-order perturbation by the $T$-operator
in the Luttinger Hamiltonian [Eq.~(\ref{HamLut})], which couples $|1,-\frac{1%
}{2}\rangle$ ($|1,\frac{1}{2}\rangle$) to $|1,-\frac{3}{2}\rangle$ ($|1,%
\frac{3}{2}\rangle$) [the second term in Eq.~(\ref{Psi1+}) or (\ref{Psi1-}%
)]. So the matrix element $\langle \Psi_{hh,+}|\hat{S}_x|\Psi_{hh,-}\rangle$
equals to
\begin{eqnarray}
\langle\Psi^{(0)}_{hh,+}|\hat{S}_x|\Psi^{(1)}_{hh,-}\rangle+\langle
\Psi^{(1)}_{hh,+}|\hat{S}_x|\Psi^{(0)}_{hh,-} \rangle=\frac{3}{4\pi^2}%
L_z^2k_-^2.  \label{Sx1+1-}
\end{eqnarray}
It is obvious that the above formula is just the off-diagonal element in $%
\tilde{S}_x$ matrix [Eq.~(\ref{Sx1})] with the first term in square bracket
of $S_1$ [Eq.~(\ref{S1kR3})] retained. This gives the quadratic-$k$
dependence of the spin orientation shown as the second term in Eq.~(\ref{Sx3}%
).

The second type comes from joint action of the $R$ in in Luttinger
Hamiltonian and the asymmetrical potential $V_a$ [See Eqs.~(\ref{Psi2+}) and
(\ref{Psi2-})]. The second-order perturbation contributes to $%
\langle\Psi_{hh,+}|\hat{S}_x|\Psi_{hh,+}\rangle$ with
\begin{eqnarray}
&&\langle\Psi^{(0)}_{hh,+}|\hat{S}_x|\Psi^{(2)}_{hh,+}\rangle +
\langle\Psi^{(2)}_{hh,+}|\hat{S}_x|\Psi^{(0)}_{hh,+}\rangle  \nonumber \\
&&+\langle\Psi^{(1)}_{hh,+}|\hat{S}_x|\Psi^{(1)}_{hh,+}\rangle  \nonumber \\
&&=-\frac{512\gamma _{2}L_{z}^{4}eFm_{0}k_{y}}{9\pi ^{6}\hbar
^{2}(3\gamma_{1}+10\gamma _{2})(\gamma _{1}-2\gamma _{2})}.  \label{Sx2+}
\end{eqnarray}
This term is just the diagonal element in Eq.~(\ref{Sx1}), which leads to
the first term in Eq.~(\ref{Sx3})
%. Since $\langle\Psi_{hh,-}|\hat{S}_x|\Psi_{hh,-}\rangle = \langle\Psi_{hh,+}|\hat{S}_x|\Psi_{hh,+}\rangle$
and is resposible for the identical spin orientation for two spin splitting
hole states in small k regime.

The spin splitting between $HH\pm$ depends on the coupling between $|1,\frac{%
3}{2}\rangle$ and $|1,-\frac{3}{2}\rangle$ through higher-order
perturbation. Different from the electron case, the direct coupling will not
cause the x-direction or y-direction spin orientation. Instead, it results
from the coupling between $|1,\frac{3}{2}\rangle$ ($|1,-\frac{3}{2}\rangle$)
and $|1,\frac{1}{2}\rangle$ ($|1,-\frac{1}{2}\rangle$). For two LH1 states,
denoted as $\Psi_{lh,\pm}$, such coupling will lead to the spin orientations
of $\Psi_{lh,+}$ opposite to $\Psi_{hh,+}$, and that of $\Psi_{lh,-}$
opposite to $\Psi_{hh,-}$. Thus the total spin orientation of the 2DHG is
conserved, though $\Psi_{hh,+}$ and $\Psi_{hh,-}$ have the same spin
orientation in the low hole density regime.

\section{Numerical Results for CISP in 2DHG}

Based on the calculated eigenstates and eigenenergies of the total
Hamiltonian (\ref{Ham1}), in this Section we will work out the spin
polarization by using the Bastin version of Kubo formula (\ref{Kubo1}) in
the relaxation time approximation. Of course the validity of such
approximation depends on the vanishing vertex correction as mentioned above.

\begin{figure}[htbp]
\begin{center}
\includegraphics[width=3.3in]
{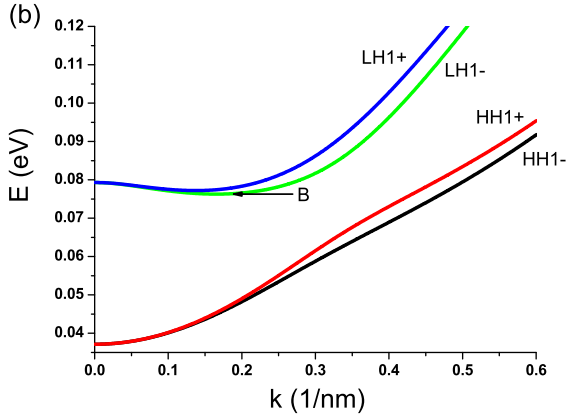}
\end{center}
\caption{ (Color online) Dispersion relation for a quantum well with
infinite barrier in an electric field. $HH1\pm$ and $LH1\pm$ denote two
lowest heavy- and light- hole subbands, respectively. The parameters for
calculation are taken as: the well width $L_z=83\mathring{A}$, the field
strength $F=50 kV/cm$, $\protect\gamma _{1}=7$ and $\protect\gamma _{2}=1.9$}
\label{Dispersion}
\end{figure}

Our numerical results with an expanded basis set of $N$ basis functions ($N$
is much larger than 8 used in last Section) shows that for a quantum well
with infinitely high potential barrier, when increasing $N$, the
eigenenergies converge to the exact solutions formulated by Huang \textit{%
et. al.}~\cite{KHuang} very quickly. For example, for the quantum well with
width $L_z=83\mathring{A}$, several lowest hole subbands obtained with $N=20$
are almost identical to the exact results. Even for $N=8$, the dispersion of
the lowest heavy and light hole subbands is in good agreement with the exact
results, demonstrating the validity of the truncation procedure in last
Section and Appendix A. Fig. ~\ref{Dispersion} plot the dispersion curves
and spin splitting of hole subbands in the quantum well in the presence of
an electric field. Due to the heavy and light hole mixture effect, the
energy minimum of the lowest light hole subband, marked by $B$ in the
Figure, deviates from the $\Gamma$-point significantly.

%As we shall show this behavior plus the spin splitting will bring about the usual

For the electric spin susceptibility, we calculate $\chi_{yx}$ only, because
$\chi_{xx}=\chi_{yy}=0$ and $\chi_{xy} = \chi_{yx}$ as indicated by Eq.~(\ref%
{chi1}).
%Since the matrix element $\langle \nu |S_{y}|\nu' \rangle \langle \nu' |v_{x}|\nu \rangle $ contains real parts only??,
After some algebra, we can divide ESS in Eq.(\ref{Kubo1}) into an
intra-subband part $\chi _{yx}^{I}$ and an inter-subband part $\chi
_{yx}^{II}$, which are expressed respectively as
\begin{eqnarray}
\chi _{yx}^{I} &=&\frac{e\hbar }{2\pi }\int \frac{d^{2}k}{(2\pi )^{2}}%
\sum_{\nu }\langle k\nu|\hat{S}_{y}|k\nu\rangle\langle k\nu|\hat{v}%
_{x}|k\nu\rangle\frac{A_{\nu }^{2}}{2},  \label{Kubo41a} \\
\chi _{yx}^{II} &=&\frac{e\hbar }{2\pi }\int \frac{d^{2}k}{(2\pi )^{2}}
\nonumber \\
&&\sum_{\nu >\nu^{\prime}}\Re (\langle k\nu|\hat{S}_{y}|k\nu^{\prime}\rangle%
\langle k\nu^{\prime}|\hat{v}_{x}|k\nu\rangle) A_{\nu }A_{\nu^{\prime}}.
\label{Kubo41b}
\end{eqnarray}
Here $\Re $ denotes the real part, and $\nu $ and $\nu^{\prime}$ stand for
the hole subband. In relaxation time approximation, the spectral function $%
A_{\nu }$ can be expressed as
\begin{equation}
A_{\nu }=\frac{2\eta }{((E-E_{\nu })^{2}+\eta ^{2})^{2}},
\end{equation}%
where $\eta =\frac{\hbar }{2\tau }$.

\begin{figure}[htbp]
\begin{center}
\includegraphics[width=3.3in]
{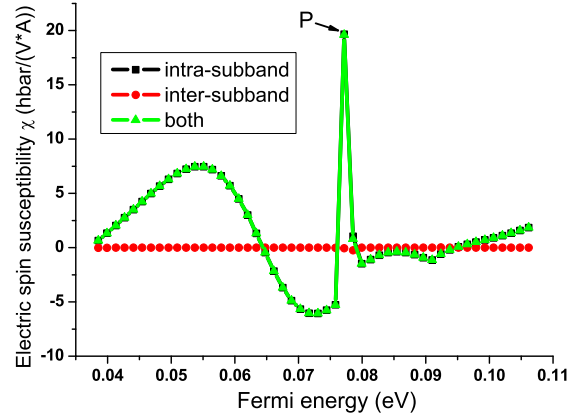}
\end{center}
\caption{ (Color online) The calculated ESS for the intra-subband term
(black square line), inter-subband term (red circle line), and their sum
(green triangle line). The scattering induced broadening $\protect\eta$ is
taken as $1.65\times10^{-5}eV $, corresponding to the relaxation time $%
\protect\tau =2\times10^{-11}s$. The spin polarization peak marked with $P$
corresponds to the energy minimum of the lowest light hole subband marked as
$B$ in Fig. \protect\ref{Dispersion}. }
\label{SpinPol}
\end{figure}

A typical curve for the CISP is plotted in Fig.~\ref{SpinPol}. The main
contribution to CISP comes from the intra-subband term, which can be
understood by Eq.~(\ref{Kubo41b}). In the limit $\eta \rightarrow 0$, the
spectral function $A_{\nu }$ tends to be the delta-function $2\pi \delta
(E-E_{\nu })$, making the inter-subband term $A_{\nu}A_{\nu^{\prime}}$ to
vanish except for an accidental degeneracy. Several features in Fig.~\ref%
{SpinPol} are worth pointing out. First, in low doping regime where only $%
HH1\pm $ states near $\Gamma $ point are occupied, spin polarization
exhibits a linear dependence on the Fermi energy. Second, with the hole
density increased, the spin polarization increases at first, then decrease
after reaching a maximum value, and even changes its sign when the hole
density is large enough. Third, when the doping is so heavy that the light
hole subband is occupied, a sharp peak for the spin polarization may be
observed as marked as $P$ in Fig. \ref{SpinPol}.

To understand these features, we turn back to Eq.~(\ref{Kubo41a}), as main
contribution to the spin polarization stems from this intra-subband term.
Based on numerical results as well as Eq.~(\ref{Sy3}), we adopt a function $%
J_{\nu }(k)$ to express the amplitude of the spin orientation associated
with the subband $\nu$, i.e.
\[
(S_{y})_{\nu \nu }=J_{\nu }(k) \cos \theta.
\]
Then, with
\[
(v_{x})_{\nu \nu }=\frac{1}{\hbar}\frac{\partial E_{\nu }}{\partial k_{x}}=%
\frac{1}{\hbar }\frac{\partial E_{\nu }(k)}{\partial k}\cos \theta,
\]
and
\[
A_{\nu }^{2}=\frac{4\pi \tau }{\hbar }\delta (E_{F}-E_{\nu }),
\]
we rewrite Eq.~(\ref{Kubo41a}) as
\begin{equation}
\chi _{yx}=\frac{e\tau }{4\pi \hbar }\sum_{\nu }k_{\nu }^{F}J_{\nu }(k_{\nu
}^{F}),  \label{ChiyxNum1}
\end{equation}%
where $k_{\nu }^{F}$ is the Fermi momentum with the hole subband $\nu $.
%This expression (\ref{ChiyxNum1}) directly relates CISP to the function $J_{\nu }(k)$ at the Fermi surface.

In the $k$-cubic Rashba model, in which only the lowest heavy hole subband $%
HH1\pm $ is concerned, up to the first-order in $\alpha $, the Fermi
momentum can be expressed as $k_{\mu }^{F}=\frac{\sqrt{ 2m_{h}E_{F}}}{\hbar }%
-\mu \frac{2\alpha m_{h}^{2}E_{F}}{\hbar ^{4}}$. Combined with Eq.(\ref{Sy3}%
),
%$J_{\mu }(k)$ contains a spin-independent part $S_{0}k$ and a spin-dependent part $\mu S_{1}k^{2}$. Substituting $J_{\mu }(k)$ and $k_{\mu }^{F}$ into Eq.(\ref{ChiyxNum1}),
we obtain
\begin{equation}
\chi _{yx}=\frac{e\tau m_{h}S_{0}E_{F}}{\pi \hbar ^{3}}+\frac{3e\tau
m_{h}^{3}\alpha S_{1}E_{F}^{2}}{\pi \hbar ^{7}}.  \label{ChiyxNum2}
\end{equation}

\begin{figure}[htbp]
\begin{center}
\includegraphics[width=3.3in]
{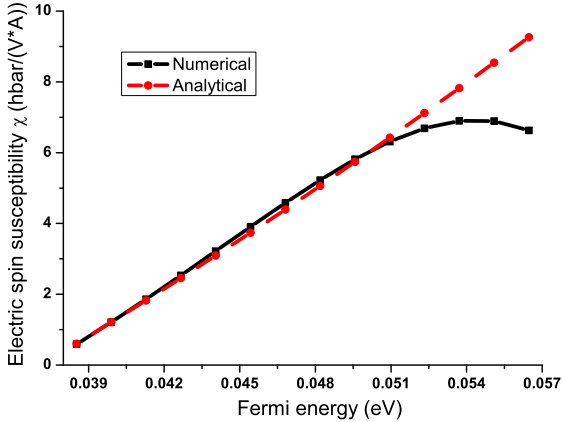}
\end{center}
\caption{ (Color online) Numerically calculated ESS as functions of Fermi
energy (black-square-line) compared with the analytical results
(red-circle-line).}
\label{NumAnaly}
\end{figure}

The first term on the right hand side of Eq.~(\ref{ChiyxNum2}), resulting
from the spin-independent part, is identical to Eq.~(\ref{ChiyxkR3}); while
the second term, proportional to $E_{F}^2$, can be safely ignored in the low
density regime. As shown in Fig.~\ref{NumAnaly}, the analytical results of
the electric spin susceptibility (Eq.~ \ref{ChiyxNum2}) agree well with the
numerical ones, demonstrating the applicability of $k$-cubic Rashba model (%
\ref{HamkR3}) in low doping regime. However, for higher hole density,
numerical results show a drop of the $\chi$ due to the heavy and light hole
mixing effect, which is certainly beyond the simple $k$-cubic Rashba model.

\begin{figure}[htbp]
\begin{center}
\includegraphics[width=3.3in]
{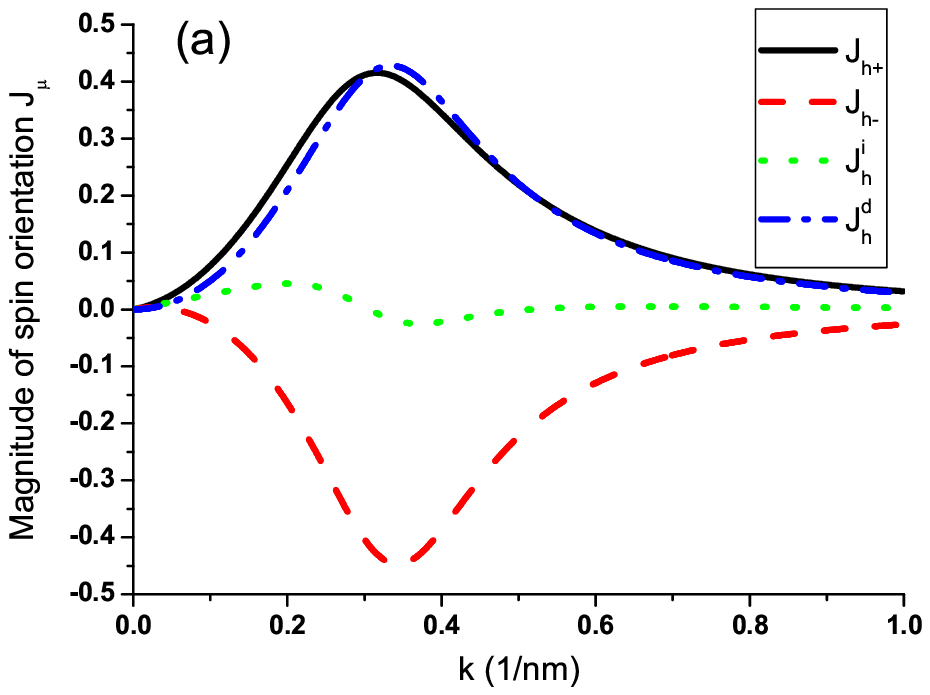} \includegraphics[width=3.3in] {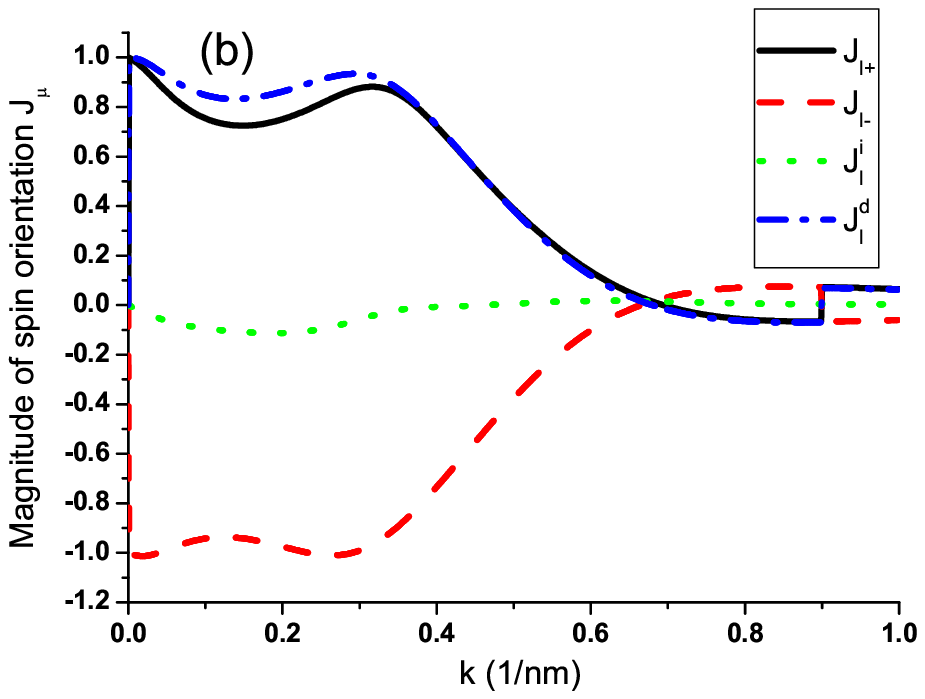}
\end{center}
\caption{ (Color online) The magnitude of the spin orientation for the
lowest heavy hole subband (a) and the lowest light hole subband (b). In (a),
the black solid line and red dashed line represent $J_{h+}$ and $J_{h-}$,
respectively; while the green dotted line and blue dashed dotted line denote
the spin- independent part $J^i_h$ and dependent part $J^d_h$, respectively.
The same notions are also applied to (b).}
\label{SyElement}
\end{figure}

For numerical results, similar to the derivation above, we may divide $%
J_{\nu } $ into a spin-dependent part and a spin-independent one, namely,
%Now instead of $\mu $, we use two indices $n$ and $\xi $ to denote the sub-bands: $n$ for heavy or light hole sub-bands quantized along z-direction and $\xi =\pm 1 $ for opposite spins $\pm \frac{3}{2}$ or $\pm \frac{1}{2}$. With the partition of
$J_{\nu \mu }=J_{\nu}^{i}+ \mu J_{\nu}^{d}$. Then the ESS can be expressed
as
\begin{equation}
\chi _{yx}=\chi _{yx}^{i}+\chi _{yx}^{d},
\end{equation}%
in which the spin- independent and dependent part respectively reads
\begin{eqnarray}
\chi _{yx}^{i} &=&\frac{e\tau }{2\pi \hbar }\sum_{\nu}J_{\nu}^{i}\frac{
k_{\nu+}^{F}+k_{\nu-}^{F}}{2},  \label{ChiyxiNum3} \\
\chi _{yx}^{d} &=&\frac{e\tau }{2\pi \hbar }\sum_{\nu}J_{\nu}^{d}\frac{%
k_{\nu+}^{F}-k_{\nu-}^{F}}{2}.  \label{ChiyxdNum3}
\end{eqnarray}%
Obviously. $\chi _{yx}^{i}$ depends on the average of Fermi wavenumbers,
while $\chi _{yx}^{d}$ depends on the Fermi wavenumber difference between
two spin-split branches. In most cases, owing to the fact that the spin
splitting is small compared with the Fermi energy, $\chi _{yx}^{i}$ will
dominate the spin polarization. In Fig. \ref{SyElement}(a), we plot the
magnitude of spin orientation associated with the subband $HH1\pm $, denoted
by $J_{h\pm }$, and the corresponding spin- dependent part, $J_{h}^{d}$, and
independent part $J_h^i$. They are related through $J_{h}^{d}=\left(
J_{h+}-J_{h-}\right) /2$ and $J_{h}^{i}=\left( J_{h+}+J_{h-}\right) /2$.
Fig.~\ref{SyElement} indicates that for most values of $k$ $J_{h}^{d}$ is
larger than $J_{h}^{i}$. Compared to the intra-subband contribution in Fig. %
\ref{SpinPol}, the spin-independent magnitude of the spin polarization $%
J_{h}^{i}$ [green-dotted line in Fig. \ref{SyElement}(a)] has similar
behavior: first increasing linearly with $k$, then decreasing with $k$
increased, and even changing the sign for larger $k$.

\begin{figure}[htbp]
\begin{center}
\includegraphics[width=3.3in]
{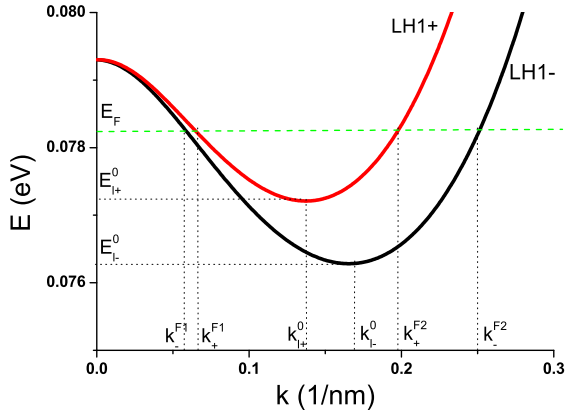}
\end{center}
\caption{ (Color online) Dispersion relation of the lowest light hole
subband $LH1\pm$. }
\label{Dispersion2}
\end{figure}

A pronounced peak of CISP may appear when the Fermi energy just crosses the
bottom of the lowest light hole subband $LH1-$. As amplified in Fig.~\ref%
{Dispersion2}, in the dispersion relation of the subband $LH1\pm$, the wave
numbers $k_{l\pm }^{0}$ corresponding the energy minimum $E_{l\pm }^{0}$
deviate from the $k=0$ point significantly. Around the energy minimum the
energy dispersion can be approximated as $E_{l\mu }(k)=E_{l\mu }^{0}+\frac{1%
}{2}\frac{\partial ^{2}E_{l\mu }(k)}{\partial k^{2}}(k-k_{l\mu }^{0})^{2}$.
Assuming the above energy dispersion and a constant magnitude of $J_{l\mu }$%
, we obtain
\begin{equation}
\chi _{yx}^{\mu}=\frac{e\tau }{2\pi \hbar }k_{l\mu }^{0}J_{l\mu }(k_{l\mu
}^{0}),  \label{ChiyxNum4}
\end{equation}%
where $k_{l\mu }^{0}\simeq (k_{\mu }^{F1}+k_{\mu }^{F2})/2$, and $k_{\pm
}^{F1}$ and $k_{\pm }^{F2}$ respectively denote two different Fermi wave
numbers for $LH1\pm$ (Fig. \ref{Dispersion2}). By Eq.~(\ref{ChiyxNum4}) and
Fig.~\ref{SyElement}(b), we can see since $J_{l+ }$ and $J_{l- }$ are large
in the absolute value but almost opposite in the sign, when $%
E_{l+}^{0}>E_{F}>E_{l-}^{0}$, a large spin polarization ${e\tau
k_{l-}^{0}J_{l-}}/(2\pi \hbar )$ is expected; on the other hand, when $%
E_{F}>E_{l+}^{0}>E_{l-}^{0}$, the contributions of $LH1\pm $ to the spin
polarization cancel each other to some extent, resulting in
\begin{equation}
\chi _{yx}=\frac{e\tau }{2\pi \hbar }%
[(k_{l-}^{0}+k_{l+}^{0})J_{l}^{i}+(k_{l+}^{0}-k_{l-}^{0})J_{l}^{d}].
\label{ChiyxNum5}
\end{equation}%
%
%
%
%where $J_{l\mu }=J_{l}^{i}+\mu J_{l}^{d}$.
As $J_{l}^{i}$ is much smaller than $J_{l}^{d}$ or $J_{l\pm }$, and $%
k_{l+}^{0}\approx k_{l-}^{0}$, both terms in Eq.~(\ref{ChiyxNum5}) are small
compared to the case when only $LH_-$ is occupied. Apparently, the peak
width depends on the spin splitting between $LH_-$ and $LH_+$.

\begin{figure}[htbp]
\begin{center}
\includegraphics[width=3in]
{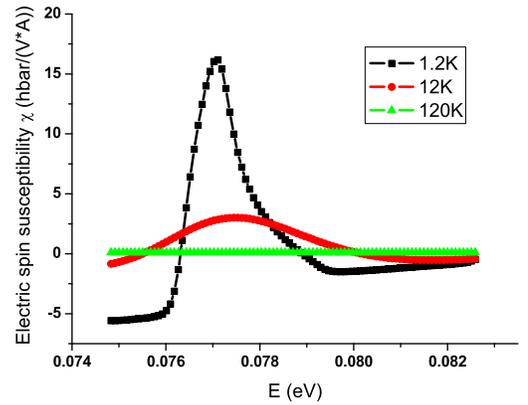}
\end{center}
\caption{ (Color online) The spin polarization peak at three different temperatures, 1.2K,
12K and 120K. }
\label{peak}
\end{figure}

The temperature dependence of the peak is plotted in Fig. \ref{peak}. Near
the polarization peak, if we only take into account $LH1\pm$, ESS is
expressed by
\begin{equation}
\chi _{yx}=\frac{e\tau }{2\pi \hbar }\sum_{\mu}f(E^0_{l\mu})k^0_{l\mu
}J_{l\mu }(k^0_{l\mu}).  \label{ChiyxNum6}
\end{equation}%
At zero temperature, the Fermi distribution function $f(E)$ becomes the
step-function $\theta(E_f-E)$, which reproduces the above analysis. At
finite temperature $T$, if we approximate $k^0_{l\mu }J_{l\mu}(k^0_{l\mu})
\simeq \mu k^0_lJ_l$, and expand the Fermi distribution function at large $%
k_BT$ as $f(E)=\frac{1}{2}(1-\frac{ E-E_F}{2k_BT})$ ( $k_B$ is Boltzmann
constant), then Eq.(\ref{ChiyxNum6}) reduces to
\begin{equation}
\chi _{yx}=\frac{e\tau k^0_{l}J_{l}}{2\pi\hbar}\frac{E_{l+}^0-E_{l-}^0}{4k_BT%
}.
\end{equation}
So ESS is proportional to the ratio of the spin splitting of the LH1
subband, $E_{l+}^0-E_{l-}^0$, to thermal energy $k_BT$. When $k_BT$ is much
larger than the spin splitting, this pronounced spin polarization peak will
smear out.

\begin{figure}[tbph]
\begin{center}
\includegraphics[width=3.3in]
{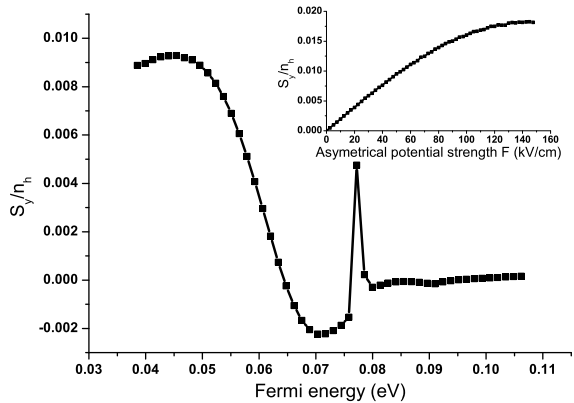} %\includegraphics[width=3.3in] {Syjx.EPS}
\end{center}
\caption{ $\frac{\langle S_{y}\rangle }{n_{h}}$ as functions of the Fermi
energy. The inset: $\frac{\langle S_{y}\rangle }{n_{h}}$ as functions of
applied field strength $F$. }
\label{SyNhjx}
\end{figure}

Now let's estimate the magnitude of the averaged CISP. In the $k$-cubic
Rashba model with an applied filed $F=50kV/cm$, Eq.(\ref{S0kR3}) gives $%
S_{0}=2.74\mathring{A}$ for $L_{z}=83\mathring{A}$ and $S_{0}=5.77\mathring{A%
}$ for $L_{z}=100\mathring{A}$. If typical relaxation time $\tau $ is taken
to be $2\times 10^{-11}s$ and an in-plane electric field strength $%
E_{0}=10V/cm$, the Fermi sphere will be shifted by $\Delta k=eE_{0}\tau
/\hbar =3\times 10^{-3}\mathring{A}^{-1}$. Substituting the above data into
Eq.(\ref{ChiyxkR3}), we obtain $\langle S_{y}\rangle /n_{h}=0.831\%$ for $%
L_{z}=83\mathring{A}$, and $\langle S_{y}\rangle /n_{h}=1.75\%$ for $%
L_{z}=100\mathring{A}$. Since $S_{0}$ is proportional to $L_{z}^{4}$, the
spin polarization is very sensitive to the thickness of quantum well. Hence,
it is preferable to detect the CISP in a thicker quantum well
experimentally. The above estimation gives the same order of magnitude for
the spin polarization observed in Silov's experiment\cite{Silov1}. In Fig.~%
\ref{SyNhjx}, we plot the averaged spin polarization $\langle S_{y}\rangle
/n_{h}$ as functions of the Fermi energy and functions of the field $F$ in
the inset. The CISP is saturated about $2\%$ when the field is enhanced.

\section{Summary}

In conclusion, we have systematically investigate the current induced spin
polarization of 2DHG in the frame of the linear response theory. We
introduce the physical quantity of the electric spin susceptibility $\chi$
to describe CISP and give its analytical expression in the simplified $k$%
-cubic Rashba model. Different from the 2DEG, the CISP of 2DHG depends
linearly on the Fermi energy. The difference of CISP between 2DHG and 2DEG
results from the different spin orientations in the subband of carriers. We
propose that $k $-cubic Rashba coefficient of 2DHG can be deduced from the
ratio of spin polarization to the current, which is independent of the
impurities or disorder effect up to the lowest order. We have also carried
out numerical calculations for the CISP. The numerical results are
consistent with the analytical one in low doping regime, which demonstrates
the applicability of $k$-cubic Rashba model. With the increase of Fermi
energy, numerical results show that the spin polarization may be suppressed
and even changes its sign. We predict and explain a pronounced spin
polarization peak when the Fermi energy crosses over the subband bottom of
the $LH_-$. We also discuss the possibility of measuring this spin
polarization peak.

\begin{acknowledgments}
This work was supported by the Research Grant Council of Hong Kong under
Grant No.: HKU 7041/07P, by the NSF of China (Grant No.10774086, 10574076),
and by the Program of Basic Research Development of China (Grant No.
2006CB921500).
\end{acknowledgments}

\begin{appendix}

\section{Derivation of the $k$-cubic Rashba Hamiltonian}

In this Appendix, we present the detailed derivation of the
$k$-cubic Rashba model by means of the perturbation method.~\cite
{Bfzhu,Shen00prb,Winkler1,Winkler2,Foreman,Foreman2,Foreman3,Habib} First we truncate
the Hilbert space of the basis wave functions (\ref{Basiswf}) into
the subspace with only the lowest eight states
$\mathcal{G}_{0}=\{|n,\lambda _{h}\rangle ,n=1,2;\lambda _{h}=\pm
\frac{3}{2},\pm \frac{1}{2}\}$. As described in the Sec. II, by
comparing the lowest HH and LH subband dispersion with the exact
solution, the accuracy of such truncation procedure has been
verified. The truncated subspace $\mathcal{G}_{0}$ can be further
cast into two sub-groups, $\mathcal{G}_{1}$ and $\mathcal{G}_{2}$.
$\mathcal{G}_{1}$ contains two lowest heavy hole states
$\{|1,3/2\rangle ,|1,-3/2\rangle \} $, while $\mathcal{G}_{2}$ keeps
the other six states, $\{|1,1/2\rangle ,|1,-1/2\rangle
,|2,3/2\rangle ,|2,-3/2\rangle $, $|2,1/2\rangle ,|2,-1/2\rangle
\}$. In this case, the Hamiltonian in the subspace $\mathcal{G}_{0}$
can be written in the form of block matrices as
\begin{equation}
H_{8\times 8}=\left(
\begin{array}{cc}
\tilde{H}_{2\times 2} & \tilde{H}_{2\times 6} \\
\tilde{H}_{6\times 2} & \tilde{H}_{6\times 6}%
\end{array}%
\right) ,  \label{AppHam881}
\end{equation}%
where
\begin{equation}
\tilde{H}_{2\times 2}=\left(
\begin{array}{cc}
P(1) & 0 \\
0 & P(1)%
\end{array}%
\right) ,
\end{equation}%
\begin{equation}
\tilde{H}_{6\times 2}=\tilde{H}_{2\times 6}^{\dag }=\left(
\begin{array}{cc}
0 & T \\
T^{\dag } & 0 \\
eFG(2,1) & 0 \\
0 & eFG(2,1)\\
R(2,1)k_+ & 0 \\
0 & -R(2,1)k_-
\end{array}%
\right) ,
\end{equation}%
and
\begin{widetext}
\begin{eqnarray}
\tilde{H}_{6\times6}=\left(\begin{array}{cccccc}Q(1)&0&R(1,2)k_+&0&eFG(1,2)&0\\0&Q(1)&0&-R(1,2)k_-&0&eFG(1,2)\\R(2,1)k_-
    &0&P(2)&0&0&T\\
    0&-R(2,1)k_+&0&P(2)&T^{\dag}&0\\eFG(2,1)&0&0&T&Q(2)&0\\0&eFG(2,1)&T^{\dag}&0&0&Q(2)\end{array}\right).
\end{eqnarray}
\end{widetext}
Here $P(n),Q(n),G(n,m),R(n,m)$ are given by
\begin{eqnarray}
P(n) &=&\frac{\hbar ^{2}}{2m_{0}}\left[ (\gamma _{1}+\gamma
_{2})k^{2}+(\gamma _{1}-2\gamma _{2})(\frac{n\pi }{L_{z}})^{2}\right] , \\
Q(n) &=&\frac{\hbar ^{2}}{2m_{0}}\left[ (\gamma _{1}-\gamma
_{2})k^{2}+(\gamma _{1}+2\gamma _{2})(\frac{n\pi }{L_{z}})^{2}\right] ,
\end{eqnarray}%
\begin{equation}
G(n,m)=\frac{4L_{z}nm((-1)^{n+m}-1)}{\pi ^{2}(m^{2}-n^{2})^{2}},
\end{equation}%
\begin{equation}
R(n,m)=-2\sqrt{3}\frac{\hbar ^{2}\gamma _{3}}{2m_{0}}\frac{%
2inm((-1)^{n+m}-1)}{L_{z}(n^{2}-m^{2})}.
\end{equation}

Our aim is to perform a transformation which decouples the groups
$\mathcal{G}_{1}$ from $\mathcal{G}_{2}$, i.e. to make the
off-diagonal part $\tilde{H}_{2\times 6}$ and $\tilde{H}_{6\times
2}$ vanish up to the first-order in $k$ and $F$. We divide the
total Hamiltonian (\ref{AppHam881}) into three parts
\begin{equation}
H_{8\times 8}=H_{0}+H_{1}+H_{2}.
\end{equation}%
The first term $H_{0}$ is the diagonal matrix elements of
$H_{8\times8}$, given by
\begin{equation}
H_{0}=\left(
\begin{array}{cc}
\tilde{H}_{2\times 2}^{(0)} & 0 \\
0 & \tilde{H}_{6\times 6}^{(0)}%
\end{array}%
\right) ,
\end{equation}%
with $\tilde{H}_{2\times 2}^{(0)}=Diag[P(1),P(1)]$ and $\tilde{H}%
_{6\times
6}^{(0)}=Diag[Q(1),Q(1),P(2),P(2),Q(2),Q(2)]$.

The second term $H_{1}$ is given by
\begin{equation}
H_{1}=\left(
\begin{array}{cc}
0 & 0 \\
0 & \tilde{H}_{6\times 6}^{(1)}%
\end{array}%
\right) ,
\end{equation}%
where %$\tilde{H}_{2\times 2}^{(1)}=\tilde{H}_{2\times 2}-\tilde{H}_{2\times
%2}^{(0)}$ and
$\tilde{H}_{6\times 6}^{(1)}=\tilde{H}_{6\times 6}-\tilde{H}%
_{6\times 6}^{(0)}$. The third term $H_{2}$ contains the non-diagonal part $%
\tilde{H}_{2\times 6}$ and $\tilde{H}_{6\times 2}$
\begin{equation}
H_{2}=\left(
\begin{array}{cc}
0 & \tilde{H}_{2\times 6} \\
\tilde{H}_{6\times 2} & 0%
\end{array}%
\right) .
\end{equation}%
There are three types of perturbation terms  in $H_1$ and $H_2$:
(1)The k-linear $R$ term couples the state $|n,\frac{3}{2}\rangle$
($|n,-\frac{3}{2}\rangle$) with $|m,\frac{1}{2}\rangle$
($|m,-\frac{1}{2}\rangle$), where $n$ and $m$ must be of opposite
parities due to the presence of $k_z=-i\partial_z$; (2) The
k-quadratic $T$ term couples $|n,\frac{3}{2}\rangle$
($|n,-\frac{3}{2}\rangle$) with $|n,-\frac{1}{2}\rangle$
($|n,\frac{1}{2}\rangle$); (3) The asymmetric potential $V_a$
couples the states with the same spin index and different
parities.

The perturbation procedure is as follows. First $H_2$ will be
eliminated by the canonical transformation as
\begin{eqnarray}
&&H_{8\times 8}^{(1)}=\exp[-U^{(1)}]H_{8\times 8}\exp[U^{(1)}]
\nonumber \\
&&=H_{8\times 8}+[H_{8\times 8},U^{(1)}]+\frac{1}{2}[[H_{8\times
8},U^{(1)}],U^{(1)}]\nonumber\\
&&+...,
\end{eqnarray}
in which $ U^{(1)}$ is chosen such that
$$H_{2}+[H_{0},U^{(1)}]=0,$$ and the matrix elements read
\begin{equation}
U_{\alpha \beta }^{(1)}=-\frac{(H_{2})_{\alpha \beta }}{E_{\alpha
}-E_{\beta }},\qquad \alpha \neq \beta,
\end{equation}%
where $E_{\alpha}$ denotes the energy of the band $\alpha$ at the $\Gamma$ point (k=0).
After the canonical transformation, the new Hamiltonian is given
by
\begin{eqnarray}
H_{8\times
8}^{(1)}=H_{0}+H_{1}+\frac{1}{2}[H_{2},U^{(1)}]+[H_{1},U^{(1)}]
\nonumber \\
 +\frac{1}{2}[[H_{1},U^{(1)}],U^{(1)}]+\cdots .
\end{eqnarray}
The $H_{0}$, $H_{1}$, $\frac{1%
}{2}[H_{2},U^{(1)}]$ and $\frac{1}{2}[[H_{1},U^{(1)}],U^{(1)}]$
have the block-diagonal form, while $[H_{1},U^{(1)}]$ is non-
block-diagonal and contains new terms first-order in $k$. So we
divide $H_{8\times 8}^{(1)}$ into three parts again
\begin{equation}
H_{8\times 8}^{(1)}=H_{0}+H_{1}^{(1)}+H_{2}^{(1)},
\end{equation}%
in which $H_{1}^{(1)}=H_{1}+\frac{1}{2}[H_{2},U^{(1)}]+\frac{1}{2}
[[H_{1},U^{(1)}],U^{(1)}]$, and $H_{2}^{(1)}=[H_{1},U^{(1)}]$. We
perform the second canonical transformation $U^{(2)}$, given by
\begin{equation}
U_{\alpha \beta }^{(2)}=-\frac{(H_{2}^{(1)})_{\alpha \beta
}}{E_{\alpha }-E_{\beta }},\qquad \alpha \neq \beta .
\end{equation}%
This makes the non-diagonal block matrix $H_{2}^{(1)}$ zero,
leading to the Hamiltonian
\begin{eqnarray}
H_{8\times 8}^{(2)}=H_{0}+H_{1}^{(1)}+\frac{1}{2}
[H_{2}^{(1)},U^{(1)}]+[H_{1}^{(1)},U^{(1)}] \nonumber \\
+\frac{1}{2}[[H_{1}^{(1)},U^{(1)}],U^{(1)}]+\cdots .
\end{eqnarray}
Now the non-block-diagonal terms of $H_{8\times 8}^{(2)}$ vanish
up to the desired order in $k$ and $F$. Finally, by mapping the
Hamiltonian $H_{8\times 8}^{(2)}$ into the lowest heavy hole
subbands, we obtain the $k$-cubic Rashba Hamiltonian
Eq.~(\ref{HamkR3}).

To obtain the corresponding spin operators in the lowest heavy
hole basis, we should apply the same canonical transformations $
U^{(1)} $ and $U^{(2)}$ to the spin operators $S_{i}$ ($i=x,y,z$).
In the 8-state subspace $\mathcal{G}_{0}$, we find that the spin
operator has the block-diagonal form
$S_{i}=Diag[S_{i}^{(1)},S_{i}^{(1)}]$ $(i=x,y,z)$, because there
are no matrix elements between the states with different
confinement quantum number $n$. Therefore $S_{i}^{(1)}$
%and $S_{i}^{(2)}$ are two
is a $4\times 4$ matrix, given respectively by
\begin{eqnarray}
S_{x}^{(1)} &=&\frac{1}{2}\left(
\begin{array}{cccc}
0 & 0 & \sqrt{3} & 0 \\
0 & 0 & 0 & \sqrt{3} \\
\sqrt{3} & 0 & 0 & 2 \\
0 & \sqrt{3} & 2 & 0%
\end{array}%
\right) ,% \\
\end{eqnarray}%
\begin{eqnarray}
S_{y}^{(1)} &=&\frac{i}{2}\left(
\begin{array}{cccc}
0 & 0 & -\sqrt{3} & 0 \\
0 & 0 & 0 & \sqrt{3} \\
\sqrt{3} & 0 & 0 & -2 \\
0 & -\sqrt{3} & 2 & 0%
\end{array}%
\right) ,% \\
\end{eqnarray}%
\begin{eqnarray}
S_{z}^{(1)} &=&\frac{1}{2}\left(
\begin{array}{cccc}
3 & 0 & 0 & 0 \\
0 & -3 & 0 & 0 \\
0 & 0 & 1 & 0 \\
0 & 0 & 0 & -1%
\end{array}%
\right) .% \\
\end{eqnarray}

Then we apply the transformations $U^{(1)}$ and $U^{(2)}$ to
the spin operators, obtaining the new spin operators $\tilde{S}%
_{i}=S_{i}+[S_{i},U^{(1)}]+[S_{i},U^{(2)}]$ as presented in Eqs. (\ref{Sx1}), (\ref%
{Sy1}) and (\ref{Sz1}).

\section{Hole Rashba term}
The hole Rashba term has recently attracted many researchers'
attentions. ~\cite{Winkler2,Bernevig,Hasegawa} The hole Rashba
term breaks the inversion symmetry, ~\cite{Bfzhu,Winkler1} and is
expressed as
\begin{eqnarray}
\hat{H}_{R}=\lambda \left(
\begin{array}{cccc}
0 & \frac{i\sqrt{3}}{2}k_{-} & 0 & 0 \\
-\frac{i\sqrt{3}}{2}k_{+} & 0 & ik_{-} & 0 \\
0 & -ik_{+} & 0 & \frac{i\sqrt{3}}{2}k_{-} \\
0 & 0 & -\frac{i\sqrt{3}}{2}k_{+} & 0%
\end{array}%
\right),
\end{eqnarray}%
where $\lambda =r_{41}^{8v8v}F$, $ r_{41}^{8v8v}$ is a parameter
as already given by Winkler for several materials,~\cite{Winkler1}
and $F$ is the field strength. If we neglect other asymmetrical
potentials and only consider the Rashba term, then the total
Hamiltonian is $\hat{H}=H_{L}+V_{c}+H_{R}$. Applying the same
perturbation procedure as in the appendix A, we find that both the
Hamiltonian and the spin operator have the identical structure to
the asymmetrical potential case, as well as the same effective
mass $ m_{h}$, $S_{1}$ and expression Eq.~(\ref{S0alphakR3}),
except for the Rashba coefficient given by
\begin{eqnarray}
\alpha =\frac{3\lambda L_{z}^{2}}{4\pi ^{2}},
\end{eqnarray}%
and the spin operator parameter
\begin{eqnarray}
S_{0}=\frac{3\lambda m_{0}L_{z}^{2}}{4\pi ^{2}\hbar ^{2}\gamma
_{2}}.
\end{eqnarray}%
The hole Rashba coefficient $\alpha $ here is proportional to
$L_{z}^{2}$, while for the asymmetrical potential case it depends
on $L_{z}^{4}$. So in most realistic quantum wells, the
contribution from the asymmetrical potential plays more important
role than the hole Rashba term, at least one or two orders of
magnitude larger. The physical reason for this may be understood
from the origin of the hole Rashba term. The more general form of
the Hamiltonian should be $\hat{H}=\hat{H}_{\mathbf{k}\cdot
\mathbf{p} }+V_{c}+eFz $, where the multi-band $\mathbf{k}\cdot
\mathbf{p}$ Hamiltonian $\hat{H}_{\mathbf{k}\cdot \mathbf{p}}$
includes not only the heavy and light hole bands, but also the
conduction band, spin split-off band and remote bands. When we
project the Hamiltonian into the subspace of the heavy and light
hole bands, the combined effects of the $eFz$ and $\mathbf{k}\cdot
\mathbf{p}$ mediated by other bands lead to the hole Rashba term,
which has much smaller influence than that coupled by the
asymmetrical potential directly. Therefore, hole Rashba term is
neglected in the present article for simplicity.

\end{appendix}

%\maketitle
%\end{CJK*}

\end{document}